\documentclass[reprint, amsmath,amssymb,prb, nofootinbib, floatfix,superscriptaddress,longbibliography]{revtex4-2}
\usepackage{empheq}
\usepackage{bbm}
\usepackage{float}
\usepackage[normalem]{ulem}
\usepackage{lipsum}
\usepackage{makecell}
\usepackage{mathtools, cuted, relsize}
\usepackage{esvect}
\usepackage{soul, xcolor}
\usepackage[caption = false]{subfig}
\usepackage{graphicx}
\usepackage{dcolumn,multirow, tabularx}
\usepackage{bm}
\usepackage{hyperref}
\usepackage{float}
\usepackage{physics}
\usepackage{slashed}
\usepackage{balance}
\usepackage[mathlines]{lineno}
\usepackage{mathtools}
\usepackage{cleveref}

\mathtoolsset{showmanualtags}
\RequirePackage{fix-cm}

\def\-{\sout}
\usepackage{titlesec}

\begin{document}

\title{Emergence of Imaginary Time Crystals in the non-Hermitian Su-Schrieffer-Heeger model}

\author{E. Slootman}
\thanks{Authors contributed equally}
\affiliation{Institute for Theoretical Physics, Utrecht University, Princetonplein 5, 3584CC Utrecht, The Netherlands \looseness=-1}%
\affiliation{MESA\textsuperscript{+} Institute for Nanotechnology, University of Twente, PO Box 217, 7500AE Enschede, The Netherlands \looseness=-1}
\author{L. Eek}
\thanks{Authors contributed equally}
 \affiliation{Institute for Theoretical Physics, Utrecht University, Princetonplein 5, 3584CC Utrecht, The Netherlands \looseness=-1}%
\author{C. Morais Smith}
\affiliation{Institute for Theoretical Physics, Utrecht University, Princetonplein 5, 3584CC Utrecht, The Netherlands \looseness=-1}%
\author{R. Arouca}
\thanks{Corresponding author: rodrigo.arouca@physics.uu.se}
\affiliation{Department of Physics and Astronomy, Uppsala University, Uppsala, Sweden}

\date{\today}

\begin{abstract}
Parity-time symmetry constrains the spectrum of non-Hermitian systems to be either real or come in complex conjugate pairs. The transition between a symmetry-preserving phase with real energies and a symmetry-broken phase with complex energies is marked by exceptional points, one of the hallmarks of non-Hermitian systems. Because of these properties, these systems are widely studied, both theoretically and experimentally. In this work, we investigate the thermodynamic properties of the gain and loss Su-Schrieffer-Heeger model for both bosons and fermions, and establish the existence of an imaginary time crystal phase, an imaginary time analogue of a time crystal. This phase occurs when there is a resonance condition between the Matsubara frequencies and the spectrum of the system, making the Green's function of the system oscillate in imaginary time with the Matsubara frequency. We show that this phase appears in the symmetry-broken region. In particular, the topological edge states of this system exhibit oscillations that are present for bosons. Finally, we discuss the applicability of our results for experiments. We examine signatures of these phases in terms of correlation functions in real time and oscillations in thermodynamic potential in inverse temperature $\beta$, and explore possible experimental platforms to realize this system.
\end{abstract}

\maketitle

\section{Introduction}\label{sec_intro}

The description of quantum systems in equilibrium by the evolution in imaginary time is an established and powerful formulation of thermodynamics. Remarkably, in this formulation, bosonic/fermionic field operators are symmetric/anti-symmetric with a period determined by the inverse temperature $\beta$. Consequently, their propagators present an infinite number of poles in the imaginary frequency domain given by the Matsubara frequencies. While these poles are essential to obtain all the thermodynamic properties, Green's functions usually are not dominated by one of them, since the energies are, in general, real or have a small imaginary part associated with a lifetime. The absence of one prevalent Matsubara frequency means that correlation functions are typically monotonic in imaginary time.  This is not the case, though, for non-Hermitian Hamiltonians, which normally exhibit a sizable imaginary part in their spectra.

Non-Hermitian systems have their time evolution described by an effective non-Hermitian operator. They appear naturally in electronic transport \cite{Datta_2005, Ochkan2024}, in the propagation of light on lossy systems in photonics \cite{Poli2015, Weimann2016, Parto2018,  Dangel2018a, Miri2019, Cherifi2023,  Slootman2024}, and as an effective description of quantum open systems \cite{Kozii2024, Roccati2022}. In the last few years, there has been a great interest in non-Hermitian systems in condensed-matter physics because the classification of topological states for non-Hermitian Hamiltonians is much richer \cite{Bergholtz2021} than for Hermitian models. This occurs because they extend the possible Altland-Zirnbauer classes \cite{Gong2018, Kawabata2019, Yang_2024} and present topological features associated with their eigenvalues \cite{Pap2018,Zhang2020a, Konig2022, Lin2023}. 

Although non-Hermitian systems are a fruitful research field, a relatively unexplored area is the thermodynamics of non-Hermitian systems. While the concept of doing thermodynamics of an intrinsically open system seems contradictory, for systems with spectra coming in complex conjugated pairs, the thermodynamic potentials are real \cite{Gardas2016}. Moreover, non-Hermitian systems exihibit remarkable critical properties \cite{Hanai2020, Arouca2020, Fruchart2021, Begg2024, ChenYe2024, MunozArboleda2024} and can even be used as quantum batteries \cite{lu2024topologicalquantumbatteries}. Since $\mathcal{PT}$-symmetric systems present complex energies in the $\mathcal{PT}$-broken phase, one expects that the correlation functions of these systems carry signatures of the poles in the Matsubara frequencies. In fact, for a generic non-interacting non-Hermitian system, the Green's function shows oscillations in imaginary time for completely imaginary energy modes \cite{Arouca2022}. Moreover, since the partition function of a non-interacting system is determined solely by its Green's functions, the thermodynamic potentials also show oscillations in $\beta$ \cite{Arouca2022}. The oscillations in both imaginary time and $\beta$ characterize the imaginary time crystal (iTC) phase, which was first conjectured by Wilczek in his seminal work on time crystals \cite{Wilczek2012} and further characterized in Refs.~\cite{Cai2020, Arouca2022}. The iTC phase is also connected to the theory of Yang-Lee of phase transitions and a special kind of critical phase called disorder points \cite{stephenson1970ising, chakrabarty2011modulation, chakrabarty2012universality, timonin2021disorder}. In Ref.~\cite{Arouca2022}, the existence of this phase in non-Hermitian quantum gases was shown for the Hatano-Nelson model \cite{Hatano1996}. However, the characterization of this phase in more general topological systems is still an open problem.

\begin{figure*}
    \centering
    \includegraphics[]{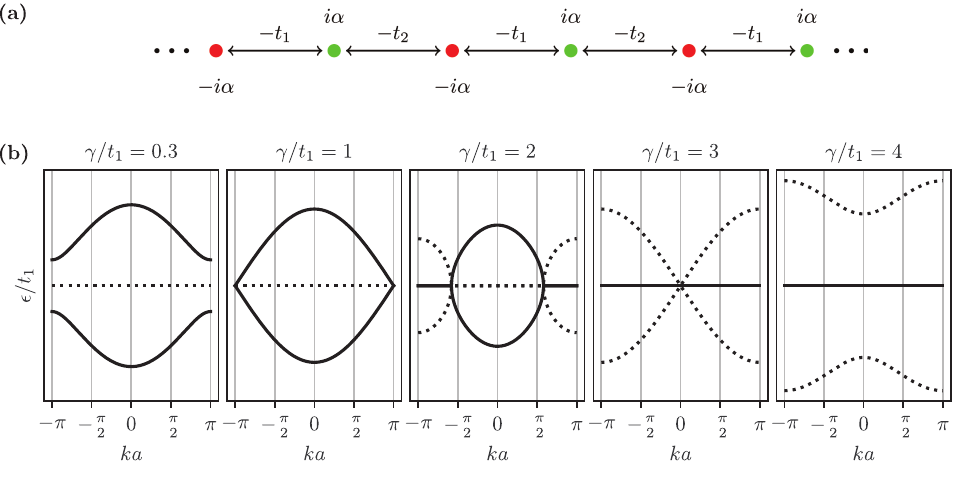}
    \caption{(a) nH-SSH lattice with alternating gain and loss. Green (red) circles represent sites on sublattices $A$ ($B$). (b) Spectrum of the SSH model with alternating gain and loss in the PBC regime, with $t_2/t_1 = 2$, for different values of $\gamma/t_1$. Solid line designates the real part of the energy, and the dashed line the imaginary part.}
    \label{fig:fig1}
\end{figure*}

In this work, we explore the existence of iTC in one of the paradigmatic parity-time($\mathcal{PT}$)-symmetric models, the gain and loss Su-Schrieffer-Heeger (SSH) model \cite{Su1979, Schomerus:13, Poli2015, Weimann2016, Parto2018, Dangel2018a, Miri2019, Slootman2024}, which is a dimerized chain with an alternating pattern of loss and gain, see Fig.~\ref{fig:fig1}(a). Due to its simplicity, the gain and loss SSH has been realized in a variety of experimental platforms . Moreover, this model presents a topological phase similar to the regular SSH model, but with complex energies when $\mathcal{PT}$-symmetry is spontaneously broken. In particular, the edge states have completely imaginary dispersion and have opposite signs. Therefore, just one of the modes survives for a long time. By analyzing the spectrum and Green's function of this model, we determine the existence of the iTC phase in this system. Moreover, we expand on the experimental signature of this phase and possible experimental platforms.

The outline of this paper is as follows. In Sec. \ref{sec_intro_iTC}, we briefly review why non-Hermitian systems present poles in their Green's functions, leading to oscillations in imaginary time. Afterwards, in Sec. \ref{sec_res}, we investigate the spectrum and pole structure of the gain and loss SSH model, showing how both properties determine the behavior in real space and imaginary time of the Green's function. We find that for a high value of loss compared to the hopping parameter, one obtains poles in finite Matsubara modes for both bosonic and fermionic systems, which lead to oscillations in both real space and imaginary time. In particular, since the topological edge states are purely imaginary, they exhibit oscillations in imaginary time for fermionic systems. The experimental signatures of this phase, both in terms of correlation functions in time and thermodynamic potentials, and an experimental realization are discussed in Sec.~\ref{sec_exp}. Finally, we present the conclusions of the work in Sec.~\ref{sec_conc}. 

\section{Imaginary time crystals in non-Hermitian systems}\label{sec_intro_iTC}

Here, we revise the thermodynamic description of non-Hermitian quantum gases and show that they can present oscillations in imaginary time \cite{Arouca2022}. To investigate the thermodynamic behavior of this model in imaginary time, we write the partition function of a generic quantum gas\cite{Stoof2008}
\begin{equation}
    \mathcal{Z} = \int D \Psi^\dagger D\Psi e^{-S[\Psi^\dagger, \Psi]},
\end{equation}
with $S[\Psi^\dagger, \Psi]$ the action, described by
\begin{align}
    S[\Psi^\dagger, \Psi] &= \int_0^{\hbar\beta}d\tau \Psi^\dagger(\tau) (\hbar \partial_\tau -\mu + H)\Psi(\tau) \notag \\
    &\equiv \int_0^{\hbar\beta} d\tau \Psi^\dagger(\tau) G^{-1}(\tau) \Psi(\tau).
\end{align}
The fields $\Psi(\tau)$ parameterize the coherent states of the system, $\mu$ is the chemical potential, and $H$ is the Hamiltonian governing the system\footnote{We distinguish between the Hamiltonian in second quantization $\mathcal{H}$, which is an operator in the Fock space (that can be written as $\hat{\mathcal{H}}$), and the Hamiltonian operator $H$ that acts in the position/inner degree of freedom space.}. The imaginary time $\tau$ is obtained by performing an analytical continuation from real to imaginary time, by substituting $t = -i \tau$. For non-interacting quantum gases, the action is quadratic, and therefore, the path integral can be performed exactly, yielding
\begin{equation}
    \mathcal{Z}_{B/F} = \det(\hbar\partial_\tau - \mu + H)^{\mp 1} = \det(G)^{\pm 1}.\label{eq:partition}
\end{equation}
We denote the partition function for bosons and fermions with subscripts B and F, respectively. All relevant thermodynamic properties of the model may be extracted from $\mathcal{Z}_{B/F}$. By performing a Fourier transformation in imaginary time, we obtain
\begin{equation}
    \widetilde{G}_n = \frac{1}{-i\hbar \omega_n - \mu +H},
    \label{eq:greens1}
\end{equation}
where $\omega_n$ are the discrete Matsubara frequencies, which arise due to the periodicity of imaginary time in $\hbar \beta$. They are defined as
\begin{equation}
    \hbar \omega_n = \frac{\pi}{\beta}n_M = \frac{\pi}{\beta} \times \begin{cases} 2n & \text{for bosons,} \\ 2n+1 & \text{for fermions,}\end{cases}
    \label{eq:def_mat}
\end{equation}
where $n \in \mathbb{Z}$ and $n_M$ are called Matsubara modes. In the remainder of this work, we will focus on systems described by non-Hermitian Hamiltonians, i.e. $H\neq H^\dagger$. Consequently, we will make use of the biorthogonal formalism. In this case, the resolution of identity is given by
\begin{equation}
    \sum_m \phi_m^R\phi_m^{L\dagger} = \mathbb{I},
\end{equation}
where $\phi_m^{R(L)}$ corresponds to right (left) eigenvectors. By inserting the biorthogonal resolution of identity into Eq.~\eqref{eq:greens1}, we obtain
\begin{equation}
    \widetilde{G}_n = \sum_m \phi_m^R \phi_m^{L\dagger} \frac{1}{-i\hbar \omega_n -\mu + \epsilon_m}.
\end{equation}
The poles of this Green's function are given by
\begin{equation}
    \epsilon_m = i\hbar\omega_n + \mu.
    \label{eq:resonance}
\end{equation}
For Hermitian systems, $\epsilon_m \in \mathbb{R}$ such that Eq.~\eqref{eq:resonance} has no solutions, except when $i\hbar\omega_n = 0$. Nevertheless, for a non-Hermitian Hamiltonian the spectrum is generally complex, $\epsilon_m \in \mathbb{C}$, such that Eq.~\eqref{eq:resonance} has solutions for non-trivial Matsubara frequencies. Furthermore, close to the pole, the Green function can be approximated by
\begin{equation}
    G(\tau)\approx\beta\sum_m\phi_m^R \phi_m^{L\dagger} \frac{e^{-i\omega_n \tau}}{\text{Re}\left(\epsilon_m-\mu\right)}.
\end{equation}
Therefore, one can conclude that the system's Green's function oscillates in imaginary time with frequencies equal to the Matsubara frequencies.

\begin{figure*}
    \centering
    \includegraphics[]{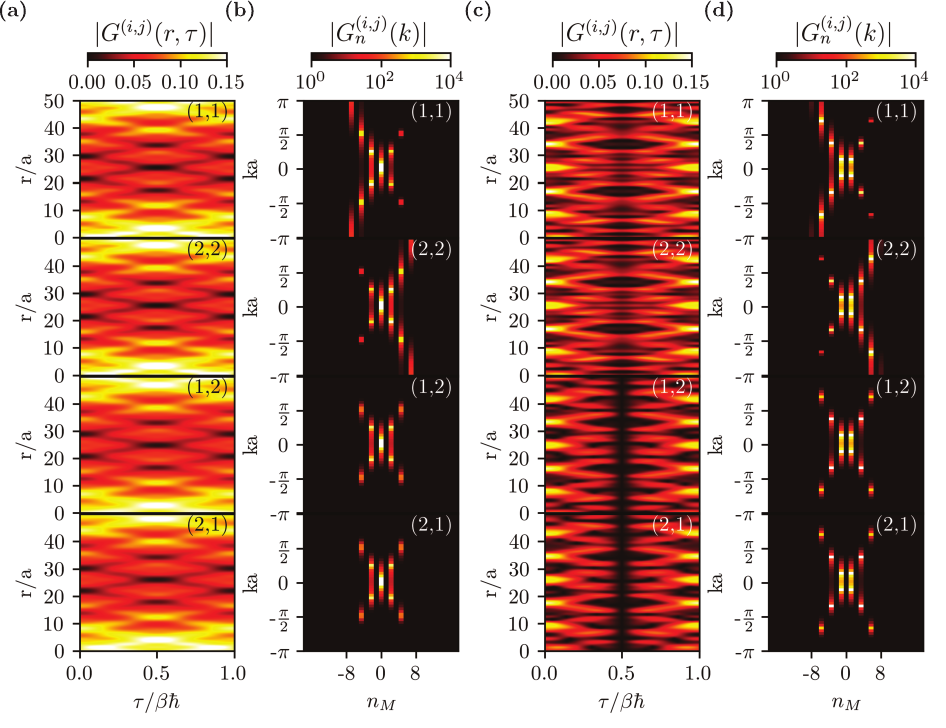}
    \caption{Green's function for the parameters $\gamma/t_1 = 3$, $t_2/t_1=2$, and $\beta=2\pi$ for bosons [(a)-(b)] and fermions [(c)-(d)]. (a) and (c) depict the Green's function in real space and imaginary time; (b) and (d) depict it as a function of momentum and Matsubara modes.}
    \label{fig:greensrt}
\end{figure*}

\begin{figure*}
    \centering
    \includegraphics[]{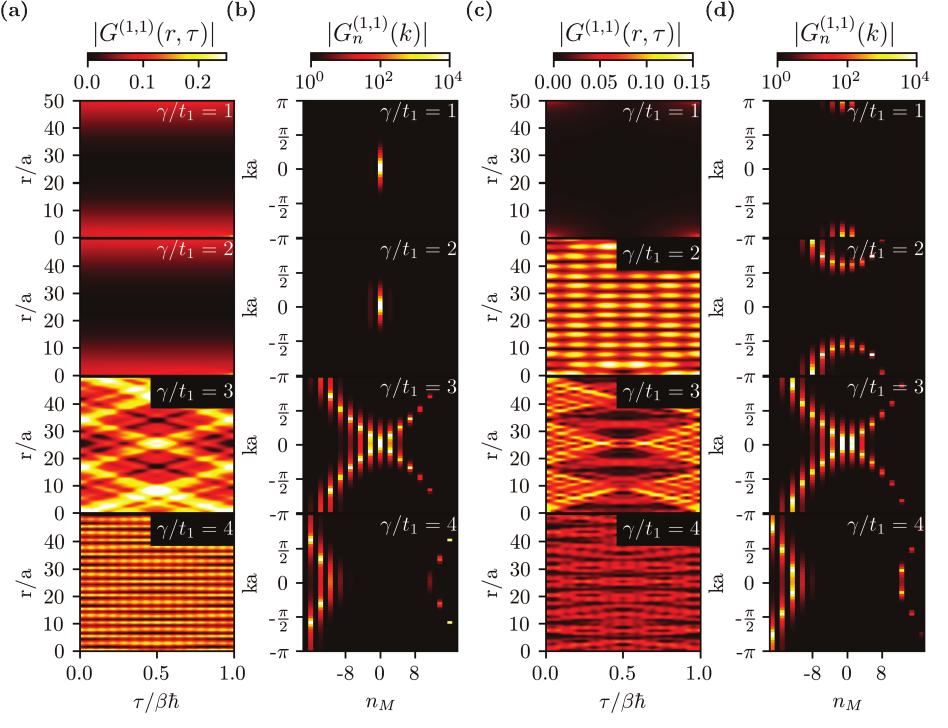}
    \caption{(a)-(d) $(1,1)$ component of the Green's function for $t_2/t_1=2$, and $\gamma/t_1$ increasing from $1$ to $4$  at a temperature of $\beta=4\pi$ for bosons [(a)-(b)] and fermions [(c)-(d)]. For all figures, the frequency of the oscillation increases as the magnitude $\gamma$ of the gain and loss is increased. However, there are no oscillations for the bosonic system, as long as it still has a partially real spectrum.}
    \label{fig:greensbig}
\end{figure*}

\section{The non-Hermitian SSH model with onsite gain/loss}\label{sec_res}

We consider a $\mathcal{PT}$-symmetric version of the non-Hermitian SSH (nH-SSH) model \cite{Schomerus:13, Poli2015, Weimann2016, Parto2018, Dangel2018a, Miri2019, Slootman2024}, where we introduce non-Hermiticity through an alternating gain and loss term $\pm i \gamma$ on the different sublattices. The Hamiltonian of this system is given by
\begin{equation}
\begin{split}
     \mathcal{H} =&\  -\sum_{i=1}^N \Big[ \left(t_1 a_i^\dagger b_i+ t_2b_i^\dagger  a_{i+1}+H.c.\right) \\
     &\ + i \gamma a_i^\dagger a^{}_i - i \gamma b_i^\dagger b^{}_i \Big],
     \label{eq:hamiltonian}
\end{split}
\end{equation}
where $N$ is the number of unit cells, $t_1$ and $t_2$ denote the intra- and inter-cell hopping, respectively, and $\gamma$ denotes the strength of on-site gain/loss. Furthermore, we assume $t_1$, $t_2$ and $\gamma$ to be real and positive. The operators $a_i$ and $b_i$ ($a_i^\dagger$ and $b_i^\dagger$) annihilate (create) a particle at site $i$ on sublattice $A$ and $B$, respectively. A schematic representation of the lattice corresponding to Eq.~\ref{eq:hamiltonian} is depicted in Fig.~\ref{fig:fig1}(a). The dispersion relation for this Hamiltonian for periodic boundary conditions (PBC) is 
\begin{equation}
    \epsilon(k) = \pm \sqrt{-\gamma^2 + t_1^2 + t_2^2 + 2t_1t_2 \cos(ka)},\label{eq:dispersion}
\end{equation}
where $a$ denotes the lattice spacing. When $t_1 \ne t_2$, the Hermitian SSH model (i.e. $\gamma =0$) exhibits a gapped spectrum. Upon introducing gain and loss ($\gamma \neq 0$), the system becomes non-Hermitian and this gap turns into a line gap. Since there is only a line gap and no point gap, the system does not display the non-Hermitian skin effect. As a consequence, the model under PBC presents the same properties as under open boundary conditions (OBC), except for the topological edge modes. In the regime where $0 \le \gamma/t_1 < t_2/t_1 - 1$, the PBC spectrum is fully real and the system has a real line gap. At $\gamma/t_1 = t_2/t_1 - 1$ the line gap is closed and when $\gamma/t_1$ is increased beyond this point, the spectrum becomes imaginary, and eventually an imaginary line gap opens up when $\gamma/t_1 \ge t_2/t_1 + 1$. This is illustrated in Fig.~\ref{fig:fig1}(b).

\subsection{Green's function for the nH-SSH model: PBC}
We now investigate the behavior of the Green's function of the nH-SSH model for PBC. Identifying the label $m$ of the energy levels in Eq.~\eqref{eq:resonance} with $\pm$ and $k$, we can insert the dispersion relation given by Eq.~\eqref{eq:dispersion} into Eq.~\eqref{eq:resonance} and use Eq.~\eqref{eq:def_mat} to obtain the resonance condition for the nH-SSH model
\begin{equation}
\begin{split}
    n_M &= \pm\frac{\beta}{\pi}\text{Im}\left[\sqrt{-\gamma^2 + t_1^2 + t_2^2 + 2t_1t_2\cos(ka)}\right],\\
    \mu &= \pm\text{Re}\left[\sqrt{-\gamma^2 + t_1^2 + t_2^2 + 2t_1t_2\cos(ka)}\right].
    \label{eq:resonancessh}
\end{split}
\end{equation}
When we consider a bosonic (fermionic) gas, the chemical potential is effectively equal to $\mu = \text{min}_k\left\{ \Re [\epsilon_-(k)]\right\}$ ($\mu = 0$ for fermions) \footnote{To avoid divergences in the numerical calculations, we will shift the chemical potential by an additional factor of $-10^{-5}$ for both bosons and fermions.}. This distinction between bosons and fermions is made because the Bose-Einstein distribution is not defined for $\Re(\epsilon-\mu) < 0$ \cite{salinas2013introduction}, while the Fermi-Dirac distribution is.
Since the spectrum of the nH-SSH model is only composed of either real or imaginary energies, see Fig.~\ref{fig:fig1} (b), for bosons the resonance condition can only be satisfied if the total spectrum is completely imaginary. This may be understood by considering the case where there are some states with negative energy: only the lowest real energy state will satisfy the condition for the chemical potential $\mu$ in Eq.~\eqref{eq:resonancessh}. Since this state has no imaginary energy, it will resonate with $n_M = 0$, and it will not show oscillations. For fermions this problem does not occur, as we take $\mu = 0$.

Since the unit cell of the nH-SSH model is composed of two sites, the Green function admits a matrix structure: $G^{(i,j)}(r,\tau)$ with $i,j \in \{1,2\}$. The $G^{(i,j)}(r,\tau)$ for bosons and fermions are shown in Figs.~\ref{fig:greensrt}(a) and \ref{fig:greensrt}(c), respectively, as a function of real space $r$ and imaginary time $\tau$ for $\gamma/t_1 = 3$, $t_2/t_1 = 2$, and $\beta = 2\pi$. For all entries of the imaginary time Green's function, a periodic pattern can be observed in the $\tau/\hbar\beta$-direction. The different frequencies that compose this periodic pattern can be understood by considering the resonance condition given by Eq.~\eqref{eq:resonancessh}. The imaginary eigenvalues are the largest at the edges of the Brillouin zone, and therefore they resonate with the largest Matsubara modes. For the chosen parameter values, we obtain $-2\sqrt{5} < n_M < 2\sqrt{5}$, i.e. only Matsubara modes $n_M = \pm 4, \pm 2, 0$ can be resonant for bosons ($n_M = \pm 3, \pm 1$ for fermions). This is confirmed by Fig.~\ref{fig:greensrt}(b) and \ref{fig:greensrt}(d), which show the Green's function in momentum space and as a function of Matsubara modes, i.e., upon performing a Fourier transformation in both the spatial and (imaginary) temporal directions. For the bosonic (fermionic) Green's function, the Matsubara modes can only take on even (odd) integer values (hence the black vertical lines in Fig.~\ref{fig:greensrt}(b) and \ref{fig:greensrt}(d) at odd (even) modes. The Green's function shows clear divergences only for the resonant Matsubara modes. These modes give rise to the periodic behavior of the Green's function in imaginary time. Furthermore, since the lossy sites are located on the first sublattice (sublattice $1$), the $G^{(1,1)}_n(k)$ Green's function will resonate with a negative Matsubara mode. Meanwhile, the gain is located on the second sublattice (sublattice $2$), and therefore $G^{(2,2)}_n(k)$ will resonate more strongly with positive Matsubara modes. The other two, $G^{(1,2)}_n(k)$ and $G_n^{(2,1)}(k)$ propagate over both sublattices and thus have a symmetric pole structure. Upon increasing $\beta$, more Matsubara modes will start to contribute, and higher frequencies will show up in the imaginary time Green's function.

The value of $\gamma$ is of utmost importance to the oscillatory behavior of $G^{(i,j)}$ in imaginary time. This is exemplified in Fig.~\ref{fig:greensbig}. Here, Figs.~\ref{fig:greensbig}(a) and \ref{fig:greensbig}(c) depict the real space and imaginary time Green's function for bosons and fermions, respectively. Similary, Figs.~\ref{fig:greensbig}(b) and \ref{fig:greensbig}(d) show the momentum space and Matsubara mode, i.e. the Fourier transformed Green's function. For $\gamma/t_1 \leq 1$, $\mathcal{PT}$-symmetry is unbroken throughout the whole Brillouin zone and, consequently, the spectrum is fully real. In this case, the only Matsubara mode that can resonate is $n_M = 0$, which does not lead to oscillations in $\tau$. For $\gamma/t_1 > 1$, part of the spectrum becomes imaginary, which allows for more Matsubara modes to satisfy the resonance condition for fermions. For bosons, oscillations are only possible for a fully imaginary spectrum, whereas for fermions, they may arise when the spectrum still has a real component. This is depicted in the second row of Fig.~\ref{fig:greensbig}(d), where oscillations occur for fermions, but not yet for bosons. If $\gamma/t_1$ increases further, the spectrum becomes fully imaginary, and divergences occur for higher $n_M$. As a consequence, imaginary time oscillations also arise for bosons. For large $\gamma$, the spectrum only contains large imaginary energies, such that only large Matsubara modes resonate. This implies high-frequency imaginary time oscillations, as depicted in the last row of Fig.~\ref{fig:greensbig} in real space.

\subsection{Green's function for the nH-SSH model: OBC}
\begin{figure}
    \centering
    \includegraphics[]{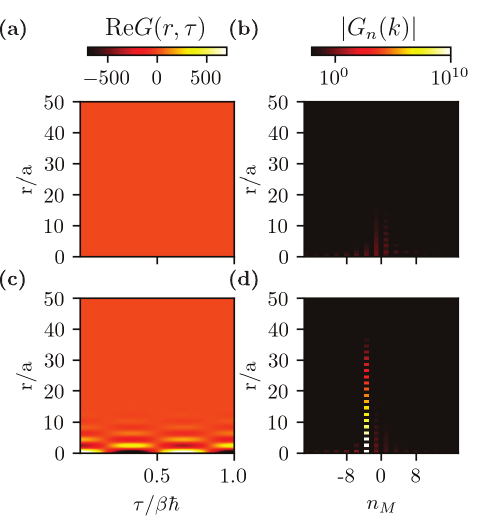}
    \caption{Green's function for OBC for fermions and the parameters $\beta = 3\pi$, $\gamma/t_1 = 1$, and $t_2/t_1 = 1/2$ [(a) and (b)] and $t_2/t_1 = 2$ [(c) and (d)]. (a) and (b) show the trivial phase of the nH-SSH model and (c) and (d) show the topological phase.}
    \label{fig:obc}
\end{figure}

We now briefly shift our attention to the model with OBC. In this case, the nH-SSH model hosts topological edge modes. While for PBC the nH-SSH Hamiltonian has a $\mathcal{PT}$-symmetric phase, when considering OBC, the presence of edge modes breaks $\mathcal{PT}$-symmetry \cite{Slootman2024}. Since the topological edge states sit at $\Re E = 0$, we focus on the fermionic case, as this allows for $\mu=0$. Figures ~\ref{fig:obc}(a)-(b) show the Green's functions for OBC in the trivial phase ($t_2/t_1 = 1/2$), while Figs.~\ref{fig:obc}(c)-(d) depict the same Green's functions for the topological phase ($t_2/t_1 = 2$). 

In the trivial phase, the model does not host topological edge modes; therefore no oscillations occur for this parameter choice, see Figs.~\ref{fig:obc}(a) and \ref{fig:obc}(b). In the topological phase, there are edge modes with $\Re \epsilon_m = 0$, satisfying the real part of Eq.~\eqref{eq:resonance}. Therefore, oscillations are present in Fig.~\ref{fig:obc}(c) because the imaginary part of the edge state resonates with the third Matsubara mode, at the chosen temperature ($\beta = 3 \pi$).

\section{Experimental signatures of an imaginary time crystal}\label{sec_exp}
Let us now comment on the signatures of a potential iTC in an experimental setting.
\begin{figure*}
    \centering
    \includegraphics[]{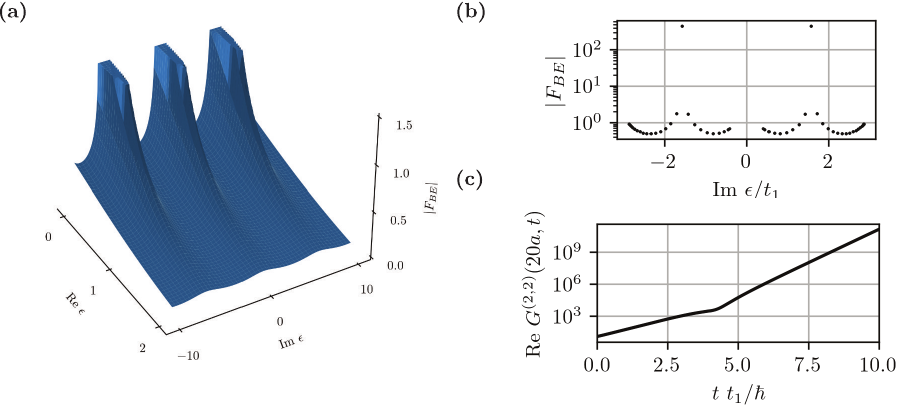}
    \caption{(a) Bose-Einstein distribution in the complex plane. The function becomes periodic for imaginary energy. (b) Absolute value of the Bose-Einstein distribution for the imaginary part of the spectrum for the model with parameters $t_2/t_1 = 2$, $\mu = 10^{-5}$, $\gamma/t_1=3.028$. The distribution diverges near $\text{Im} \left\{\epsilon/t_1 \right\}\approx \pm \pi/2$. (c) The resulting real-time behaviour of the corresponding bosonic Green's function at a fixed position in space ($r = 20$).}
    \label{fig:realgreens}
\end{figure*}

\subsection{Real-time behavior of Green's functions}
Up to this point, we have mostly considered the Green's functions in imaginary time. It would also be useful to consider a description of the model in real time. Upon Wick rotating, the Green's function in real time and close to its poles can be approximated by \cite{Lothman2021}
\begin{equation}
    G(t) = \sum_m \phi_m^R \phi_m^{L\dagger} F_{\text{BE}/\text{FD}}(\epsilon_m - \mu, \beta) \exp(-i\epsilon_m t).
    \label{eq:realtgreens}
\end{equation}
Here, $F_{\text{BE}/\text{FD}}(\epsilon_m-\mu, \beta) = \left\{ \exp[\beta (\epsilon_m-\mu)] \mp 1\right\}^{-1}$ is the Bose-Einstein/Fermi-Dirac distribution \footnote{Here, we introduced $\epsilon_m$, which is defined to be an element of the spectrum given by Eq.\eqref{eq:dispersion}, i.e. $\epsilon_m \in \left\{ \epsilon_\pm (k) \quad | \quad  k \in [0, 2\pi) \right\}$.}. As opposed to the imaginary time case, the Green's function exponentially decays/increases with the energies in real time. These decaying energies are multiplied by the distribution function. For now, let us consider the Bose-Einstein distribution. The shape of the Bose-Einstein distribution for complex energies is given in Fig.~\ref{fig:realgreens}(a). As expected, it decays as a function of real energy. Contrarily, on the imaginary axis it oscillates. Indeed, the imaginary energies at which the Bose-Einstein distribution diverges correspond to Matsubara modes given by 
\begin{equation}
    \beta \epsilon_m = 2\pi i \quad \text{for} \quad \Re\left( \epsilon_m - \mu\right) = 0. \label{eq:BEcond}
\end{equation}
Correspondingly, when the magnitude of the Bose-Einstein distribution is plotted as a function of the imaginary part of the spectrum $\{ \epsilon_m \}$, see Fig.~\ref{fig:realgreens}(b), the modes $\epsilon_m$ satisfying Eq.~\eqref{eq:BEcond}\footnote{We note that since the number of $k$-points is finite in our simulations, we do not see a singularity for all values of $\gamma$. Therefore, we choose this value with $\gamma/t_1\approx 3$, where the peak is visible.} are clearly visible. For $\beta = 4$, these occur at $\epsilon = \pm i\pi/2 $. When considering the real time behaviour of the Green's function, the short-time regime is dominated by these modes.  This can be observed in Fig.~\ref{fig:realgreens}(c), which shows the real-time Green's function at a fixed position in real space on a logarithmic scale. Initially, the Green's function increases at a slope corresponding to the largest resonating Matsubara frequency: $\pi/2 \approx 1.57$.  However, after a certain time, the exponential in Eq.~\eqref{eq:realtgreens} starts to outweigh the Bose-Einstein distribution. Thereafter, the Green's function has a slope corresponding to the largest imaginary eigenvalue. 

We remark that although we are showing Green's functions, in a non-interacting system, all observables are readily obtained by knowing $G$ and using Wick's theorem. Therefore, the characteristic increase in real time as a function of temperature should also be manifest in observables. 

\subsection{Thermodynamic signatures} 
\begin{figure*}
    \centering
    \includegraphics[]{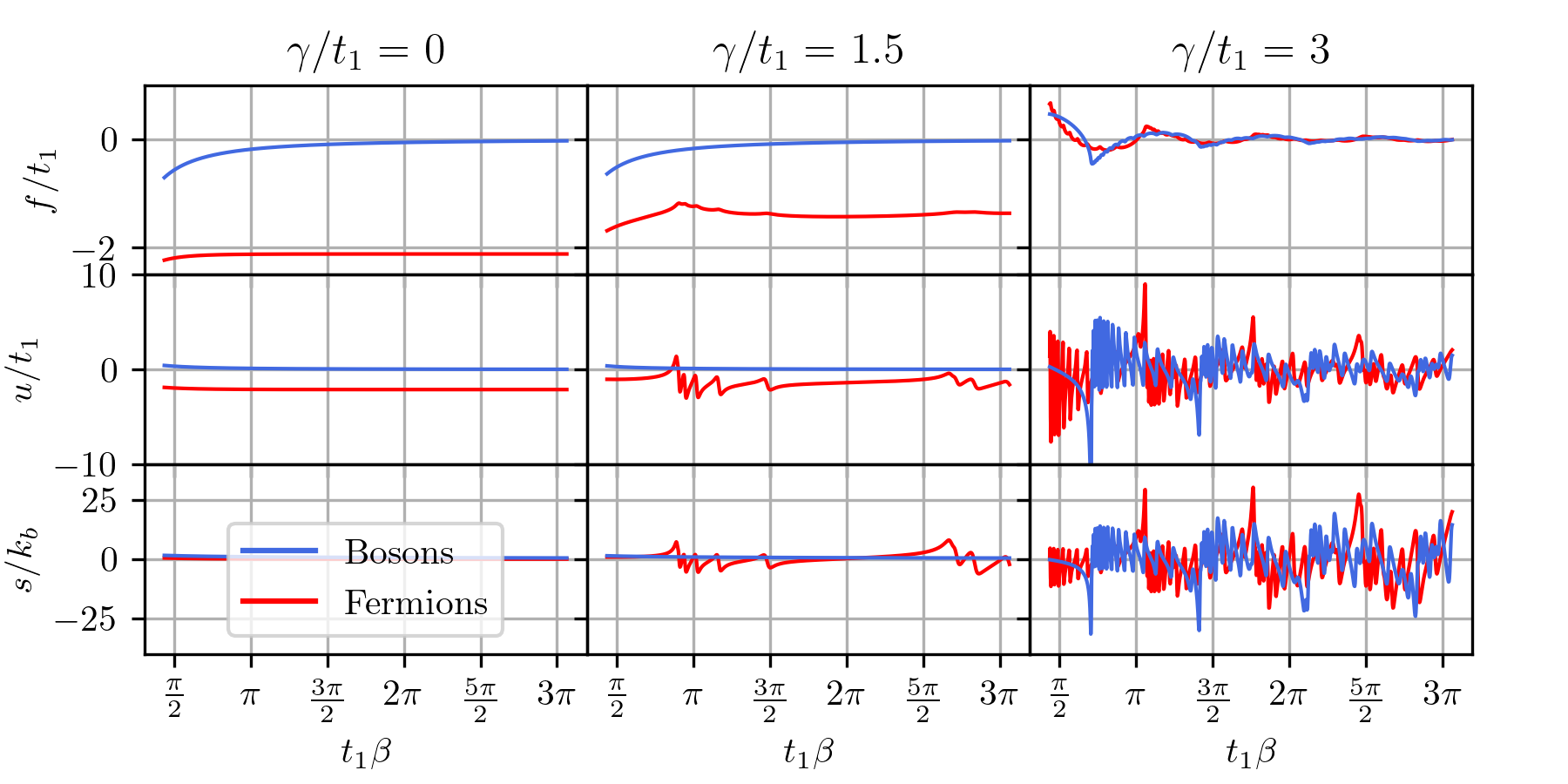}
    \caption{Free energy (top row), internal energy (middle row), and entropy (bottom row) for the model for three values of $\gamma/t_1$ for bosons (blue) and fermions (red). For the calculation of the thermodynamic quantities, we have taken the offset in $\mu$ to be $10^{-3}$.}
    \label{fig:thermo}
\end{figure*}

The most prominent signatures of an iTC phase can be observed in thermodynamic quantities, such as the free energy $F$ and entropy $S$. By evaluating Eq.\eqref{eq:partition}, we obtain
\begin{equation}
    \mathcal{Z}_{\text{B}/\text{F}}= \prod_m \left[1\mp e^{\beta (\epsilon_m - \mu)}\right]^{\mp 1},
\end{equation}
which shows oscillations in temperature for complex energies, due to the imaginary argument of the exponential. This behavior is inherited by the thermodynamic quantities that get derived from the partition function. For example, the internal energy $U$, free energy $\mathcal{F}$, and entropy $S$, respectively:
\begin{align}
    U_{\text{B}/\text{F}} &\equiv -\frac{\partial \log \mathcal{Z}_{\text{B}/\text{F}}}{\partial \beta} = \sum_m \epsilon_m F_{\text{BE}/\text{FD}}(\epsilon_m-\mu,\beta),\notag \\
    \mathcal{F}_{\text{B}/\text{F}} &\equiv -\frac{1}{\beta}\log \mathcal{Z}_{\text{B}/\text{F}} = \pm \frac{1}{\beta} \sum_m \log \left[] 1 \mp e^{-\beta(\epsilon_m-\mu)} \right], \notag\\
    \frac{S_{\text{B}/\text{F}}}{k_B} &\equiv \beta^2 \frac{\partial \mathcal{F}_{\text{B}/\text{F}}}{\partial \beta} =-\beta \mathcal{F}_{\text{B}/\text{F}} + \beta U_{\text{B}/\text{F}}.
\end{align}
The behavior of these quantities (normalized by the system size) is plotted in Fig.~\ref{fig:thermo} for bosons and fermions. Here, the free energy, the internal energy, and the entropy are given as a function of energy for three values of $\gamma/t_1$. In the case where there is no loss or gain (left column), these quantities behave as expected. For finite $\gamma/t_1$, the thermodynamic quantities start to exhibit periodic behavior in $\beta$. For small $\gamma$, the real part of the spectrum damps the periodic behavior. However, when the imaginary component of the spectrum becomes larger, the oscillations become more apparent and start to dominate (right column).

\subsection{Possible experimental platforms}
While in condensed-matter systems it is difficult to engineer non-Hermitian terms to obtain an effective nH-SSH model, one could, in principle, use metamaterials to obtain this phase. In photonic waveguides, the evolution operator in the presence of losses is described by an effective non-Hermitian Hamiltonian. Further, changing the distance between the waveguides, different hopping terms can be engineered, allowing for the realization of this model \cite{Weimann2016, Slootman2024}. In a similar way, different hoppings inside and between unit cells can be achieved by modulation of lasers in ultracold atoms \cite{Atala2013, RevModPhys.91.015005}, while loss can be engineered by using intense pulses to remove electrons from the system \cite{Benary_2022}. In addition, in cold atoms, the temperature of the system can be well adjusted, and there is a possibility of realizing both bosonic and fermionic versions of this model. A challenge is the introduction of gain with the same level of loss, but in principle, one can obtain a passive $\mathcal{PT}$-symmetric version of this model, see, for instance, Ref.~\cite{Slootman2024}. However, measuring the thermodynamic potentials with precision in temperature to see the oscillations can be challenging. An alternative would be to measure the response of a system to a quench, which depends on the integrated Green's function in imaginary time \cite{PhysRevB.84.224303}, and could provide an indirect signature of this phase.

\section{Conclusion}\label{sec_conc}

In this work, we establish the existence of an imaginary time crystal phase in the gain and loss SSH model. We show how the change in the dispersion relation for different values of the gain/loss parameter $\gamma$ reflects the change of the pole structure of the Green's function in momentum and Matsubara modes. When the system is in the $\mathcal{PT}$-broken phase, we obtain oscillations in real space and imaginary time for both fermionic and bosonic systems. Furthermore, we observe that there is a clear oscillation in imaginary time of the edge states. We show how the short-time behavior of the Green's function can be directly associated with these modes and that oscillations in the thermodynamic potentials with inverse of temperature provide a way to experimentally characterize this phase. 

For open quantum systems, there are usually quantum effects beyond a non-Hermitian description. For instance, the Liouvillian operator is, in general, a non-Hermitian operator, but also contains a part that cannot be encoded in a Hamiltonian evolution. Moreover, the thermodynamic description of non-Hermitian systems is based on the spectrum coming in complex conjugated pairs \cite{Gardas2016}, which is usually not the situation seen in experimental platforms, where loss is more prevalent. Because of these points, a very interesting outlook is understanding how this effect is manifest for more complete approaches to open systems, like the Lindbladian or Keldysh formalism.

\subsection*{Acknowledgments}
We thank A. Hemmerich for valuable input for the Possible Experimental Realization section of our paper. In addition, we thank W Cherifi, M. Bourennane, Chris Toebes, and J. Klärs for discussions about implementing this model at finite temperature in photonics. Further, we thank D. F. Munoz-Arboleda, D. S. Quevedo, E. Bergholtz, and M. Bäcklund for related collaborations. LE and CMS acknowledge the research program ``Materials for the Quantum Age'' (QuMat) for financial support. This program (registration number 024.005.006) is part of the Gravitation program financed by the Dutch Ministry of Education, Culture and Science (OCW). RA acknowledges financial support from the Knut and Alice Wallenberg Foundation through the Wallenberg Academy Fellows program KAW 2019.0309 and project grant KAW 2019.0068. 

\bibliography{main.bib}

\end{document}